\documentclass[12pt,English,urlcolor=black,linkcolor=black]{article}

\usepackage[dvips]{graphicx}
\usepackage{epsfig}
\usepackage{amsmath,amsfonts,amssymb,amsthm}
\usepackage{mathrsfs,mathtools}
\usepackage{verbatim}
\usepackage{psfrag}
\usepackage{bm}
\usepackage{bbm}
\usepackage[utf8]{inputenc}
\usepackage[square,comma,sort&compress,numbers]{natbib}
\usepackage{xcolor}
\usepackage{slashed}
\usepackage{upgreek}
\usepackage{enumitem}
\usepackage{float}

\usepackage[colorlinks,citecolor=black]{hyperref}

\usepackage{epsf,epsfig}
\usepackage{graphics}
\usepackage{subcaption}

\newcommand\blfootnote[1]{%
  \begingroup
  \renewcommand\thefootnote{}\footnote{#1}%
  \addtocounter{footnote}{-1}%
  \endgroup
}

\def\d{\mathrm{d}}

\def\vec{\mathbf}

\usepackage[errorstop]{feynmp}
\begin{document}
\thispagestyle{empty}

\begin{flushright}
{
\small
CP3-21-53\\
TUM-HEP-1365-21
}
\end{flushright}

\vspace{0.4cm}

\begin{center}
\Large\bf\boldmath
Bubble wall velocities in local equilibrium
\unboldmath
\end{center}

\vspace{-0.2cm}

\begin{center}
{Wen-Yuan Ai,$^{*1}$\blfootnote{$^*$wenyuan.ai@uclouvain.be} Bj\"{o}rn Garbrecht,$^{\dagger 2}$\blfootnote{$^\dagger$garbrecht@tum.de} and Carlos Tamarit$^{\ddagger 2}$\blfootnote{$^\ddagger$carlos.tamarit@tum.de}\\
\vskip0.4cm
{\it $^1$Centre for Cosmology, Particle Physics and Phenomenology,\\ Université catholique de Louvain, Louvain-la-Neuve B-1348, Belgium}\\

{\it $^2$Physik-Department T70, James-Franck-Stra\ss e,\\ Technische Universit\"{a}t M\"{u}nchen, 85748 Garching, Germany}\\
\vskip1.4cm}

\end{center}

\begin{abstract}

It is commonly expected that a friction force on the bubble wall in a first-order phase transition can only arise from a departure from thermal equilibrium in the plasma. Recently however, it was argued that an effective friction, scaling as $\gamma_w^2$ (with $\gamma_w$ being the Lorentz factor for the bubble wall velocity), persists in local equilibrium. This was derived assuming constant plasma temperature and velocity throughout the wall. On the other hand, it is known that, at the leading order in derivatives, the plasma in local equilibrium only contributes a correction to the zero-temperature potential in the equation of motion of the background scalar field. For a constant plasma temperature, the equation of motion is then completely analogous to the vacuum case, the only change being a modified potential, and thus no friction should appear. We resolve these apparent
contradictions in the calculations and their interpretation and show
that the recently proposed effective friction in local equilibrium
originates from inhomogeneous temperature distributions, such that the $\gamma_w^2$-scaling of the effective force is violated. Further, we propose a new matching condition for the hydrodynamic quantities in the plasma valid in local equilibrium and tied to local entropy conservation. With this added constraint, bubble velocities in local equilibrium can be determined once the parameters in the equation of state are fixed, where we use the bag equation in order to illustrate this point. We find that there is a critical value of the transition strength $\alpha_{\rm crit}$ such that bubble walls run away for $\alpha>\alpha_{\rm crit}$.

\end{abstract}

\newpage
\tableofcontents

\section{Introduction}
The early Universe may have undergone different phase transitions.  Of particular interest are first-order phase transitions, which could explain the matter-antimatter asymmetry through the mechanism of electroweak baryogenesis~\cite{Kuzmin:1985mm,Shaposhnikov:1987tw,Morrissey:2012db,Garbrecht:2018mrp}. The dramatic collision processes after bubble nucleation could also result in a stochastic background of gravitational waves (GWs)~\cite{Witten:1984rs,Kosowsky:1991ua,Kosowsky:1992vn,Kamionkowski:1993fg,Hindmarsh:2013xza}.\footnote{For recent reviews, see Refs.~\cite{Binetruy:2012ze,Caprini:2015zlo,Cai:2017cbj,Weir:2017wfa,Caprini:2018mtu,Mazumdar:2018dfl,Caprini:2019egz,Hindmarsh:2020hop}.} The recent detection of GWs~\cite{Abbott:2016blz} thus brings the new and promising opportunity to test particle physics models using GW interferometers. For example, the approval of the Laser Interferometer Space Antenna (LISA) project~\cite{Audley:2017drz} has thus triggered renewed interest in first-order phase transitions in the early Universe. Besides LISA, many other space-based GW detectors have been proposed, such as Big Bang Observer (BBO)~\cite{Corbin:2005ny}, Deci-hertz Interferometer Gravitational wave Observatory (DECIGO)~\cite{Kawamura:2011zz}, Taiji~\cite{Gong:2014mca}, and TianQin~\cite{Luo:2015ght}.

For descriptions of the various phenomena linked to cosmological first-order phase transitions, a key parameter is the bubble wall velocity. For instance, the power spectrum in gravitational waves produced by first-order phase transitions is generally an increasing function of the wall velocity, while electroweak baryogenesis usually requires a slow bubble motion to allow particles to diffuse from the bubble wall back into the plasma (see however \cite{Cline:2020jre}). Estimates of the bubble velocity are usually based on kinetic theory \cite{Dine:1992wr,Liu:1992tn,Moore:1995ua,Moore:1995si}, considerations on fluctuation-dissipation relations~\cite{Khlebnikov:1992bx,Arnold:1993wc} or non-equilibrium quantum field theory \cite{Konstandin:2014zta} and suggest a velocity-dependent friction force due to deviations from local equilibrium near the bubble wall. To study the bubble wall growth, usually an effective friction term  proportional to a phenomenological coefficient $\eta$ is added to the equation of motion of the scalar field~\cite{Ignatius:1993qn,Kurki-Suonio:1996gkq,Espinosa:2010hh}. The friction coefficient can then be fixed by matching with Boltzmann equations derived using kinetic theory or non-equilibrium quantum field theory~\cite{Megevand:2009ut,Megevand:2009gh,Huber:2011aa,Huber:2013kj}.

Recently there has been renewed interest in the bubble wall dynamics~\cite{Mancha:2020fzw,Hoeche:2020rsg,Friedlander:2020tnq,Vanvlasselaer:2020niz,Balaji:2020yrx,Cai:2020djd,Wang:2020zlf,Bigazzi:2021ucw,Lewicki:2021pgr,Gouttenoire:2021kjv,Dorsch:2021nje,DeCurtis:2022hlx}. In Ref.~\cite{Mancha:2020fzw} it was proposed that there is an effective friction force on the bubble wall, scaling as $\gamma_w^2$ where $\gamma_w$ is the Lorentz factor of the bubble wall velocity, even if the plasma is in local thermal equilibrium. This is in contradiction with what is expected generally~\cite{Turok:1992jp} and would prohibit runaway or ultrarelativistic bubbles~\cite{Bodeker:2009qy,Bodeker:2017cim,Hoeche:2020rsg} even when neglecting the friction due to deviations from local equilibrium. The effective friction in local equilibrium has been further analysed in Ref.~\cite{Balaji:2020yrx}, in which the friction effect was explicitly demonstrated in deflagration and detonation examples, while some deviations from the expression of the force in Ref.~\cite{Mancha:2020fzw} were noted. The fact that a constant velocity bubble expansion does not necessarily require non-equilibrium friction effects could be seen already in the results of Refs.~\cite{Ignatius:1993qn,Moore:1995si,Kurki-Suonio:1996gkq,Espinosa:2010hh}, while the first dedicated analysis of bubble expansion in the absence of non-equilibrium effects was made in Ref.~\cite{Konstandin:2010dm}, where it was first shown that hydrodynamic heating effects in deflagrations could reduce the force driving the expansion of the bubbles. As argued in Ref.~\cite{Balaji:2020yrx}, this hydrodynamic effect is equivalent to the friction effect of Ref.~\cite{Mancha:2020fzw}, and not only does it apply for deflagrations but also to detonations. Since processes in local thermal equilibrium are reversible, the resulting force can be characterized as non-dissipative.

The result in Ref.~\cite{Mancha:2020fzw} deserves however a closer check. While it can be shown that the effective friction in local equilibrium can be derived from the matching conditions of the fluid hydrodynamic quantities across the bubble wall~\cite{Balaji:2020yrx}, the specific formula proposed in Ref.~\cite{Mancha:2020fzw} is obtained by assuming constant fluid temperature and velocity across the bubble wall. This assumption immediately leads to an inconsistency. For a constant fluid temperature, the equation of motion of the background scalar field is completely analogous to the vacuum case, albeit with a modified potential, and no effective friction can arise from it. Indeed, as we will show in this work, a non-constant fluid temperature across the bubble wall is a necessary condition for the backreaction force to exist in local equilibrium. As a result, it can be seen that the $\gamma_w^2$ scaling of the force is an artificial result of the inappropriate assumption used in Ref.~\cite{Mancha:2020fzw}. 

In this work, we carry out a systematic study of bubble expansion in local equilibrium. Besides resolving the contradictions mentioned above, we show that in local equilibrium there is a new matching condition for the fluid hydrodynamic quantities due to local entropy conservation. As far as we know, this matching condition has not been studied in the literature so far. With this new condition, using a ``bag model'' parametrization of the fluid density and pressure away from the wall, which applies generally when admitting temperature-dependent parameters, we are able to determine the bubble wall velocity in local equilibrium for fixed values of the bag parameters (in particular the phase transition strength $\alpha_N$), for both the detonation and deflagration fluid motion modes. As we will show, solutions with a steadily moving bubble wall in local equilibrium are restricted to a narrow parameter space, requiring $\alpha_N\leq \alpha_{\rm crit}$ where $\alpha_{\rm crit}$ is typically a small number. For larger phase transition strengths, the bubble wall runs away. Regarding the scaling of the backreaction force with the wall velocity, we find that for deflagrations it grows with velocity, as predicted in Ref.~\cite{Mancha:2020fzw}, but remains below the quantitative estimates of the former reference, as was noted before in Ref.~\cite{Balaji:2020yrx}. On the other hand,  for detonations the force actually decreases with the wall velocity.

The remainder of the paper is organized as follows. In the next section we review the analysis of steady bubble walls in the general case and in the case of local equilibrium. We pay particular attention to clarifying the origin of the recently proposed effective friction on the bubble wall in local equilibrium. As a byproduct that turns out to be of central relevance, we notice a new matching condition in local equilibrium. In Section~\ref{sec:bubble_vs}, using the new condition, we build a closed system of equations for the fluid hydrodynamic quantities, allowing in particular to determine the bubble wall velocity for given values of the phase transition strength in the bag parameterization. We solve the coupled algebraic equations numerically. In Section~\ref{sec:microphysics} we explain how to make contact between the bag parametrization and particle physics models, and provide explicit detonation examples. We present our conclusions in Section~\ref{sec:Conc}. Throughout the paper, whenever this is appropriate, we use the terms ``plasma'' and ``fluid'' interchangeably. We use the metric signature $(+,-,-,-)$.

\section{Analysis on steady bubble walls}

\subsection{Total force on a bubble wall}
\label{subsec:force}

Following Refs.~\cite{Moore:1995ua,Moore:1995si}, the equation of motion of a scalar field averaged over a finite density state becomes
 \begin{align}\label{eq:eom}
  \Box\phi+\frac{\partial V(\phi)}{\partial\phi}+\sum_i\frac{\d m^2_i(\phi)}{\d\phi}\int \frac{\d^3{\bf p}}{(2\pi)^32E_i}\,f_i(p,x)=0\,,
 \end{align}
 where the index $i$ runs over the particles in the plasma, $f_i(p,x)$ are the particle distribution functions, and $E_i=\sqrt{\vec{p}^2_i+m_i^2}$ the particle energies.
 Separating the equilibrium and non-equilibrium contributions by writing
 \begin{align}
  f_i(p,x)=f^{\rm eq}_i(p,x)+\delta f_i(p,x)\,,
 \end{align}
it can be seen that the equilibrium contributions are related to the finite-temperature corrections to the effective potential, and so one can write 
 \begin{align}\label{eq:eomdeltaf}
  \Box\phi+\frac{\partial V_{\rm eff}(\phi,T)}{\partial\phi}+\sum_i\frac{\d m^2_i(\phi)}{\d\phi}\int \frac{\d^3{\bf p}}{(2\pi)^32E_i}\,\delta f_i(p,x)=0\,,
 \end{align}
 where $V_{\rm eff}(\phi,T)$ is the potential including finite-temperature corrections.
Assuming a planar wall with constant velocity propagating in the $z$-direction, going to the wall frame, multiplying the above equation by $\d\phi/\d z$ and integrating over $z$, one gets
 \begin{align}\label{eq:friction}
  \int \d z\frac{\d\phi}{\d z}\left(\Box\phi+\frac{\partial V_{\rm eff}(\phi,T)}{\partial\phi}+\sum_i\frac{\d m^2_i(\phi)}{\d\phi}\int \frac{\d^3{\bf p}}{(2\pi)^32E_i}\,\delta f_i(p,x)\right)=0\,.
 \end{align}
The first term is zero, because it is the integral of a total derivative of a quantity that vanishes for large $|z|$ where the scalar field is constant. Admitting for a nontrivial profile in the temperature  $T$, the second term gives
\begin{align}
 \int \d z \frac{\d\phi}{\d z}\frac{\partial V_{\rm eff}(\phi,T)}{\partial\phi}= \int \d z \left(\frac{\d V_{\rm eff}}{\d z}-\frac{\partial V_{\rm eff}}{\partial T}\frac{\d T}{\d z}\right)=\Delta V_{\rm eff}-\int \d z \frac{\partial V_{\rm eff}}{\partial T}\frac{\d T}{\d z}\,.
\end{align}
Substituting this into Eq.~\eqref{eq:friction} leads to
 \begin{align}
  \Delta V_{\rm eff} = \int \d z \frac{\partial V_{\rm eff}}{\partial T}\frac{\d T}{\d z}-\sum_i \int \d z \frac{\d m_i^2(\phi)}{\d z}\,\int \frac{\d^3{\bf p}}{(2\pi)^32E_i}\,\delta f_i(p,x)\,.
 \end{align}
Due to the assumption of a stationary expansion, the above equation can be interpreted as a balance of forces: on the left the driving force due to the total  change of free-energy density across the wall, on the right the total backreaction force per unit area,
\begin{align}\label{eq:Ffr}
 \frac{F_{\rm back}}{A}=&\,\int \d z \frac{\partial V_{\rm eff}}{\partial T}\frac{\d T}{\d z}-\sum_i \int \d z\frac{\d\phi}{\d z} \frac{\d m_i^2(\phi)}{\d\phi}\,\int \frac{\d^3{\bf p}}{(2\pi)^32E_i}\,\delta f_i(p,x)\,.
\end{align}

In the literature it is sometimes assumed that the second term in this backreaction force must not vanish in order to obtain a bubble expansion at constant velocity~\cite{Moore:1995ua,Moore:1995si}. However, the first term also yields a nonvanishing contribution to the backreaction force if the temperature is not constant across the bubble wall. The second term originates from $\delta f_i$ and therefore corresponds to an out-of-equilibrium effect, so that we refer it as a dissipative friction force. The contribution to the backreaction force that remains in local equilibrium (with $\delta f_i=0$) was considered for the first time in Ref.~\cite{Ignatius:1993qn}. It has also been studied subsequently in Ref.~\cite{Espinosa:2010hh,Konstandin:2010dm}, but it was incorporated as a modification of the driving force. This force is actually the “friction force” in
local equilibrium proposed in Ref.~\cite{Mancha:2020fzw} and also studied in Ref.~\cite{Balaji:2020yrx}. As we shall see, $F_{\rm back}$ for $\delta f=0$ has a different sign for detonations and for deflagrations. The same change
of sign occurs for $F_{\rm pressure}$ defined in Eq.~\eqref{eq:Fpressure} below. Therefore, one shall keep in mind that the roles of these forces as “driving force” and
“friction” can be inverted depending on the case although we do not name them case by case.

\subsection{Backreaction force in a perfect fluid}
\label{sec:localeq}

In this section, we derive expressions for the backreaction force that apply when the plasma behaves as a perfect fluid, which is not required to be in local thermodynamical equilibrium. When the latter holds in addition, one recovers the backreaction force that was proposed in Ref.~\cite{Mancha:2020fzw} and further discussed in Ref.~\cite{Balaji:2020yrx}. We will follow the argumentation of the latter reference.
The total
energy-momentum tensor reads
\begin{align}
    T^{\mu\nu}=T^{\mu\nu}_\phi+T^{\mu\nu}_f\,, 
\end{align}
where 
\begin{subequations}
\begin{align}
    T^{\mu\nu}_\phi&=(\partial^\mu\phi)\partial^\nu\phi-g^{\mu\nu}\left(\frac{1}{2}(\partial\phi)^2-V(\phi)\right)\, ,\\
   \label{eq:Tmunuf} T^{\mu\nu}_f&=(\rho_{f}+p_{f})u^\mu u^\nu-p_{f} g^{\mu\nu}\, ,
\end{align}
\end{subequations}
are the energy-momentum tensors for the background scalar field and the fluid, respectively. Here $V(\phi)$ is the zero-temperature potential for the scalar field and $\rho_f$ and $p_f$ are the energy density and pressure for the fluid, respectively.\footnote{We reserve the notation $\rho$, $p$ for quantities that include zero-temperature effects, as this will simplify the matching conditions in Section~\ref{sec:bubble_vs}.} To be general, $\rho_f$ and $p_f$ may include non-equilibrium effects. When the background field can be treated adiabatically, the fluid pressure can be written as
\begin{align}\label{eq:pdeltaV}
    p_{f}(\phi,T)=- \Delta V_T(\phi,T)+\delta p_f\,,
\end{align}
where $\Delta V_T(\phi,T)$ designates the finite-temperature corrections to the scalar potential, while $\delta p_f$ encodes the possible non-equilibrium effects.

The conservation of energy-momentum gives 
\begin{align}
\label{eq:em-conser}
    \nabla_\mu T^{\mu\nu}=0\, .
\end{align}
Below we assume that the Hubble rate is much smaller than the other rates of interest and thus we work in Minkowski spacetime. 

The quantities $u^\mu$ and $T$ are generally spacetime dependent; $u^\mu=u^\mu(x)$, $T=T(x)$. Now we study a steady state where the bubble wall moves with a constant velocity $v_w$. Working in the rest frame of the wall,  with coordinates $(t,x,y,z)$ all of the time derivatives in Eq.~\eqref{eq:em-conser} vanish.\footnote{Note that, $\rho$, $p$ and $T$ are Lorentz scalar.} Taking the direction of the bubble wall expansion in the fluid frame as the positive $z$-direction (corresponding to a planar approximation), one can write $u^\mu(z)=\gamma(z)(1,0,0,-v(z))$.\footnote{The minus sign in front of $v(z)$ is introduced 
to ensure that $v(z)$ is non-negative.} Then the $\nu=0$ and $\nu=3$ components of Eq.~\eqref{eq:em-conser} give, respectively,
\begin{subequations}
\begin{align}
\label{eq:em-con-1}
    &\omega\gamma^2 v={\rm const}\, ,\\
    \label{eq:em-con-2}
    &\omega\gamma^2 v^2+\frac{1}{2}(\partial_z\phi(z))^2+p={\rm const}\, ,
\end{align}
\end{subequations}
where 
\begin{align}
\label{eq:P}
    p=p_{f}-V(\phi)= -V_{\rm eff}(\phi,T)+\delta p_f\,,
\end{align}
(see Eq.~\eqref{eq:pdeltaV}), while $\omega=\rho+p$ with
\begin{align}
\label{eq:Rho}
  \rho=\rho_{f}+V(\phi)\, .  
\end{align} 
Note that $\omega=\rho_f+p_f$ can be identified with the enthalpy of the plasma.
Here we have combined the
energy density and pressure of the fluid with the tree-level potential energy. This notation differs from that in Ref.~\cite{Balaji:2020yrx} and leads to a pressure and density that do not vanish at zero temperature, but the advantage is that in terms of $\rho$ and $p$ the matching condition for fluid quantities across the bubble wall corresponding to Eq.~\eqref{eq:em-con-2} take the form that appears more commonly in the literature, given in Eqs.~\eqref{eq:junctionAB} below. Note that with this convention, the fluid energy-momentum tensor does not take the usual form but reads
\begin{align}
    T_f^{\mu\nu}=(\rho+p)u^\mu u^\nu -(p+V(\phi))g^{\mu\nu}\,.
\end{align}
On the other hand, the total energy-momentum tensor in a constant scalar background (as applies far from the bubble wall), will have the form of Eq.~\eqref{eq:Tmunuf} with $\rho_f$, $p_f$ substituted by $\rho,p$.
We have also defined the $\phi$-dependent effective potential $V_{\rm eff}(\phi,T)=V(\phi)+\Delta V_T(\phi,T)$.
Integrating Eq.~\eqref{eq:em-con-2} over $z$ and noticing that $\partial_z\phi(z)=0$ away from the bubble wall, one gets
\begin{align}
\label{eq:dri-back}
   -\Delta p=\Delta\{\omega\gamma^2v^2\}\, ,
\end{align}
where $\Delta$ indicates the difference between the quantities in front of the wall and behind the wall. To have a clear interpretation of Eq.~\eqref{eq:dri-back}, let us for the moment consider the case of constant $T$ such that $-\Delta p=V_{\rm eff}(\phi_s,T)-V_{\rm eff}(\phi_b,T)-\Delta \delta p_f$ where $\phi_s$, $\phi_b$ are the field values of the symmetric and broken phases, respectively. Thus $-\Delta p+\Delta \delta p_f=\Delta V_{\rm eff}$ is fixed for a given $T$ and shall be viewed as a fixed driving force due to the pressure difference, as was also argued in Section~\ref{subsec:force}. In a steady state, such a driving force should be balanced by a backreaction force. We thus identify 
\begin{align}
\label{eq:inte}
    \frac{F_{\rm pressure}}{A}\equiv\Delta V_{\rm eff}\, ,\qquad \frac{F_{\rm back}}{A}=\Delta\{\omega\gamma^2v^2 +\delta p_f\}=\Delta\{(\gamma^2-1)\omega+\delta p_f\}\,,
\end{align}
which generalizes the derivation of Ref.~\cite{Balaji:2020yrx} beyond local equilibrium ($\delta p_f=0$).
The former backreaction force can also be derived from the general result of Eq.~\eqref{eq:Ffr}. Taking the $z$-derivative of Eq.~\eqref{eq:em-con-2}, using Eq.~\eqref{eq:P} to write 
\begin{align}
 \frac{\d p}{\d z}=-\frac{\d\phi}{\d z}\,\frac{\partial V_{\rm eff}}{\partial\phi}-\frac{\d T}{\d z}\,\frac{\partial V_{\rm eff}}{\partial T}+\frac{\d\delta p_f}{\d z}
\end{align}
and making use of the scalar equation of motion \eqref{eq:eomdeltaf}, gives
\begin{align}
 \frac{\partial V_{\rm eff}}{\partial T}\frac{\d T}{\d z}-\sum_i \frac{\d\phi}{\d z} \frac{\d m_i^2(\phi)}{\d\phi}\,\int \frac{\d^3{\bf p}}{(2\pi)^32E_i}\,\delta f_i(p,x)\,=\frac{\d}{\d z} (\omega \gamma^2 v^2+\delta p_f),
\end{align}
so that the backreaction force of Eq.~\eqref{eq:Ffr} indeed matches 
Eq.~\eqref{eq:inte}. If $\delta p_f$ relaxes to zero away from the wall, it would not contribute to the backreaction force.

The expression of Eq.~\eqref{eq:inte} is valid in general, as long as the plasma behaves as a perfect fluid with an energy-momentum tensor as in Eq.~\eqref{eq:Tmunuf}. In local equilibrium, $\delta p_f=0$ and one can use thermodynamical identities to express the enthalpy through the entropy density $s$, which is itself related to the pressure:
\begin{equation}
\label{eq:Thermo}
\omega=\rho_f+p_f = T s, \quad s=\frac{\partial p_f}{\partial T}\,.
\end{equation}
It then follows that the backreaction force can be related to the entropy density,
\begin{align}
\label{eq:backforce-equi}
\frac{F_{\rm back}}{A}= \Delta\{(\gamma^2-1)Ts\}\,\quad {\rm in\ local\ equilibrium}\,.
\end{align}

The balance of forces in Eq.~\eqref{eq:inte} can be understood as capturing the following physical effects: latent heat release due to the phase transition, which drives the expansion of bubbles, plus momentum transfer and heating effects across the fluid, which can compensate the driving force and lead to an expansion at constant velocity. In the remainder of the paper, we will restrict ourselves to local equilibrium.

\subsection{A contradiction and its resolution}

Assuming constant $T$ and constant $v=v_w$ in Eq.~\eqref{eq:backforce-equi}, one recovers the friction force of Ref.~\cite{Mancha:2020fzw}
\begin{align}
 \label{eq:force}
    \frac{F_{\rm back}}{A}=(\gamma_w^2-1)T\Delta s\, .
\end{align}
This expression for the force grows with the velocity and thus suggests that runaway behaviors should be excluded. 
In realistic cases one cannot necessarily assume a constant temperature and velocity for the fluid, so that one has to solve the differential equations for $T(x)$ and $v(x)$~\cite{Ignatius:1993qn,Konstandin:2010dm,Balaji:2020yrx}. Actually, if $T$ and $v$ are simultaneously constant, then $\Delta s$ cannot be nonvanishing, as can be seen from Eq.~\eqref{eq:em-con-1}. Nevertheless, the authors in Ref.~\cite{Mancha:2020fzw} argue that one could use the assumption of constant temperature as an approximation when the active particles, which exert a significant force on the bubble wall, are much less in number than the passive particles, whose mass remains approximately constant or zero across the bubble wall. The latter constitute a large heat reservoir and absorb the heat from the active particles. If the heat capacity of the reservoir is large enough, the temperature then does not change much. In the following, we shall show that this approximation might be misleading and that the force $F_{\rm back}$ of Eq.~\eqref{eq:backforce-equi} actually vanishes if the fluid temperature $T$ is constant across the bubble wall.

The former can be seen clearly when noting that in local equilibrium there is a contradiction between the assumption of constant temperature and the existence of a friction force. The equation of motion~\eqref{eq:eom} for the scalar field in local equilibrium simply reads
\begin{align}
\label{eq:eom2}
    \Box\phi+\frac{\partial}{\partial\phi}\,V_{\rm eff}(\phi,T)=0\, .
\end{align}
For constant $T$, however, the above equation is completely analogous to the vacuum case, albeit with a modified potential, and no effective friction can arise from it. This then contradicts the statement that there could be an effective friction for constant temperature in local equilibrium. This can be seen explicitly from the general result for the total backreaction force in Eq.~\eqref{eq:Ffr}. In local equilibrium ($\delta f_i=0$), the backreaction force requires nonzero $dT/dz$ and thus vanishes if the temperature is constant. We will next show this explicitly for the force of Eq.~\eqref{eq:backforce-equi} by exploiting identities valid in local equilibrium.

In local equilibrium, it is known that the entropy is conserved locally (see, e.g., Ref.~\cite{Hindmarsh:2020hop}),
\begin{align}
\label{eq:entropy-divergence}
    \partial_\mu S^{\mu}\equiv \partial_\mu (su^\mu)=0\, .
\end{align}
The former identity can be derived from the covariant equations for the plasma that follow after accounting for the equation of motion of the scalar field. Contracting the plasma equations with the fluid 4-velocity, and using the thermodynamical identities of Eq.~\eqref{eq:Thermo} that link $\omega$ and $p$ in equilibrium, one recovers Eq.~\eqref{eq:entropy-divergence} \cite{Ignatius:1993qn,Balaji:2020yrx}. Equation~\eqref{eq:entropy-divergence}  also leads to the conservation of the total entropy $S=\int \d^3x \gamma s$ for a fluid that remains at rest at infinity \cite{Balaji:2020yrx}.
Now it can be seen that the assumption of a constant velocity of the fluid automatically leads to vanishing $\Delta s$. Consider again a steady state in the rest frame of a planar bubble wall, with $s=s(z)$. Equation~\eqref{eq:entropy-divergence} reads then
\begin{align}
  \partial_z(s(z)\gamma(z)v(z))=0\, .
\end{align}
If $v(z)$ is constant, then we are left with $\partial_z s(z)=0$, i.e., $\Delta s=0$. 
Actually, when there is no local entropy production, we have 
\begin{align}
\label{eq:sgammav}
    s(z)\gamma(z) v(z)=c_1={\rm const}
\end{align}
and Eq.~\eqref{eq:dri-back} gives 
\begin{align}
\label{eq:back-force}
    \frac{F_{\rm back}}{A}=c_1\Delta\{\gamma v T\}\quad {\rm in\ local\ equilibrium}\, .
\end{align}
The above formula immediately tells us that if $T$ and $v$ are simultaneously constant, the backreaction force is vanishing in local equilibrium. 

What if we assume a constant $T$ only? Does Eq.~\eqref{eq:back-force} indicate however a nonvanishing friction? From Eqs.~\eqref{eq:em-con-1} and~\eqref{eq:sgammav}, we have 
\begin{align}
\label{eq:gammaT}
    \gamma(z)T(z)=c_2={\rm const}\, .
\end{align} 
Therefore, in local equilibrium constant $T$ across the bubble wall leads to constant $v$ across the bubble wall.  Hence from Eq.~\eqref{eq:back-force} we see that {\it the temperature distribution must not be constant in order to incur a non-dissipative backreaction force $F_{\rm back}$ on the bubble wall}, i.e., there must be a local heating process across the bubble wall. This conclusion is consistent with the equation of motion~\eqref{eq:eom2} for the scalar field. This heating effect corresponds to the hydrodynamic obstruction of Ref.~\cite{Konstandin:2010dm}, and can be understood from the conservation of the total entropy \cite{Balaji:2020yrx}. The entropy density is dominated by the contributions from light degrees of freedom, while the phase transition makes some particles heavy. Hence, the reduction of the entropy associated with particles becoming heavier has to be compensated with an entropy increase in some regions of the plasma caused by a heating effect.

\section{Bubble velocities in local equilibrium\label{sec:bubble_vs}}

The identity of Eq.~\eqref{eq:gammaT} represents a novel constraint in local equilibrium whose consequences for the allowed bubble velocities have not been explored in the literature yet. In this section we investigate how the additional constraint impacts the usual analysis of bubble expansion modes, which relies on imposing energy-momentum conservation across the bubble wall and, if necessary, across shock fronts in which some fluid parameters develop discontinuities \cite{Steinhardt:1981ct,Gyulassy:1983rq,Kajantie:1986hq,Enqvist:1991xw,Liu:1992tn,Huet:1992ex,Ignatius:1993qn,Espinosa:2010hh}. We note that we assume from the beginning a system that remains in local equilibrium. This requires efficient scatterings of the plasma particles on the bubble interface. For sufficiently large wall speeds, the local equilibrium and thus Eq.~\eqref{eq:gammaT} could be violated and non-equilibrium effects must be taken into account. Our results thus serve only as an upper bound for the bubble wall velocities when non-equilibrium effects have been neglected. A more complete and realistic analysis of the bubble wall velocities is left for future work.

Far from the wall, the scalar background becomes a constant, such that its contribution to the total energy-momentum tensor can be absorbed as a modification to the fluid's pressure. As usually done in the literature, we will assume a ``bag model'' parameterization for the fluid plus background scalar field,  in  which the total energy-momentum tensor has the form of a perfect fluid,
\begin{align}\label{eq:Tnograds}
 T_{\rm total}^{\mu\nu}=(\rho+p)u^\mu u^\nu -p g^{\mu\nu},
\end{align}
while the pressure and density are assumed to adopt the form
\begin{subequations}\label{eq:bagprho}
\begin{align}
    &\rho=aT^4+\epsilon\,,\\
    & p=\frac{1}{3}aT^4-\epsilon\,.
\end{align}
\end{subequations}
While typically $a$ and $\epsilon$ are treated as constants in a given scalar background---which corresponds to an ultrarelativistic limit in which only the leading thermal corrections are accounted for---admitting a temperature and background dependence in $a,\epsilon$ allows to capture arbitrary dispersion relations for the plasma for a given  background, as then one can trade arbitrary $\rho(T,\phi),p(T,\phi)$ for arbitrary $a(T,\phi),\epsilon(T,\phi)$. Hence in the following discussion $a,\epsilon$ will not be required to be constants.

As our definitions of $\rho,p$ in Section~\ref{sec:localeq} have been chosen to lead to a total energy-momentum tensor like Eq.~\eqref{eq:Tnograds} in the limit of vanishing field gradients, we can reuse the equations from Section~\ref{sec:localeq}, implementing energy-momentum conservation by evaluating these on different sides of the bubble wall, where field gradients can be dropped. Denoting the quantities in front of/behind the bubble wall with a subscript ``+/-", Eqs.~\eqref{eq:em-con-1} and~\eqref{eq:em-con-2} give the following matching conditions \cite{Steinhardt:1981ct}:
\begin{subequations}
\label{eq:junctionAB}
\begin{align}
    &\omega_+\gamma_+^2v_+=\omega_-\gamma_-^2v_-\, ,\label{eq:conditionA}\\
    &\omega_+\gamma_+^2v_+^2+p_+=\omega_-\gamma_-^2v_-^2+p_-\, .\label{eq:conditionB}
\end{align}
\end{subequations}
The above quantities with the subscript ``+/-" are defined in the rest frame of the bubble wall. Note that Eq.~\eqref{eq:inte} does not provide any new information than the above matching conditions. Any solutions to Eqs.~\eqref{eq:junctionAB} automatically satisfy the balance between the driving force and the backreaction force. Actually, the above matching conditions only fix the relation between $v_+$ and $v_-$ but cannot determine the bubble wall velocity. To see this more physically, we substitute Eq.~\eqref{eq:em-con-1} into Eq.~\eqref{eq:em-con-2} (or substitute Eq.~\eqref{eq:conditionA} into Eq.~\eqref{eq:conditionB}) and obtain
\begin{align}
\label{eq:backforce2}
    \frac{F_{\rm back}}{A}=c\Delta v\, ,
\end{align}
where $c=c_1 c_2=\omega\gamma^2 v$ (see Eqs.~\eqref{eq:sgammav} and \eqref{eq:gammaT}). 
The velocity $v$ entering the parameter $c$ can be related to the bubble wall velocity $v_w$ (see below). Nonetheless, the backreaction force is not fixed by the bubble wall velocity but rather depends on the velocity difference between the fluid in front of and behind the bubble wall. Moreover, $F_{\rm back}$ can be either positive or negative, depending on the sign of $v_+-v_-$.

Equations~\eqref{eq:conditionA} and~\eqref{eq:conditionB} give
\begin{align}
    v_+v_-=\frac{p_+-p_-}{\rho_+-\rho_-}\,,\quad \frac{v_+}{v_-}=\frac{\rho_-+p_+}{\rho_++p_-}\,.
\end{align}
After some algebra, one obtains
\begin{subequations}
\label{eq:v+v-}
\begin{align}
    v_+ v_-&=\frac{1-(1-3\alpha_+)r}{3-3(1+\alpha_+)r}\, ,\\
    \frac{v_+}{v_-}&=\frac{3+(1-3\alpha_+)r}{1+3(1+\alpha_+)r}\, , \label{eq:v+/v-}
\end{align}
\end{subequations}
where 
\begin{align}
\label{eq:alpha&r}
    \alpha_+\equiv \frac{4\Delta\theta}{3\omega_+}\ , \qquad r\equiv \frac{\omega_+}{\omega_-}\,, 
\end{align}
with $\theta=(\rho-3p)/4$ being the so-called trace anomaly. If one uses Eq.~\eqref{eq:bagprho} the quantities $\alpha_+$, $r$ read
\begin{align}
\label{eq:alpha&r2}
    \alpha_+\equiv \frac{\Delta\epsilon}{a_+T_+^4}\ , \qquad r\equiv \frac{a_+ T_+^4}{a_-T_-^4}\ ,
\end{align}
with $\Delta\epsilon\equiv\epsilon_+-\epsilon_-$. From Eqs.~\eqref{eq:v+v-}, one obtains~\cite{Steinhardt:1981ct}
\begin{align}
\label{eq:relation-v+v-}
    v_+=\frac{1}{6 \left(1+\alpha _+\right) v_-}\left[1+3 v_-^2\pm\sqrt{1+6 \left(6 \alpha _+^2+4 \alpha _+-1\right) v_-^2+9 v_-^4}\right]\,.
\end{align}
Together with physical requirements from the fluid velocity profile away from the bubble wall, $\{v_+,v_-\}$ can be classified into three regimes~\cite{Espinosa:2010hh}. The first one is the so-called {\it detonation} in which $v_w=v_+>v_-\geq c_s$, where $c_s$ is the sound speed\footnote{In the bag model with constant $a$ and $\epsilon$, $c^2_s=\partial_T p/\partial_T\rho=1/3$, and we neglect the position-dependence of $c_s$ in the general case.} and $v_+$ belongs to the positive sign branch of the above formula. In this case, there is a rarefaction behind the bubble wall while the fluid in front of the wall is unperturbed. The second one is {\it deflagration} in which $c_s>v_w=v_->v_+$ and $v_+$ resides in the negative sign branch. The situation now is opposite to the detonations; there is a shock wave in front of the wall but the fluid behind the wall is unperturbed. The last one is called {\it hybrid} in which the fluid velocity is a superposition of the first two kinds and the wall velocity $v_w$ is not identified with either $v_+$ or $v_-$. For simplicity, in this work we focus on the first two typical classes. 

It would be also useful to express $v_-$ in terms of $v_+$, and we obtain from Eq.~\eqref{eq:relation-v+v-}
\begin{align}
\label{eq:v-1}
    v_-=\frac{1}{6v_+}\left[1-3 \alpha _++3 \left(1+\alpha _+\right) v_+^2\pm\sqrt{\left(1-3 \alpha _++3 \left(1+\alpha _+\right) v_+^2\right){}^2-12 v_+^2}\right]\,.
\end{align}
As we mentioned, the backreaction force $F_{\rm back}$ does not depend only on the bubble wall velocity. Actually, given one boundary condition $T_+$ (which is given by $T_{\rm nuc}$ for detonations), Eqs.~\eqref{eq:conditionA} and~\eqref{eq:conditionB} provide two relations between the three unknown quantities $\{T_-,v_+,v_-\}$, and thus cannot eliminate all the uncertainties. Below we will study the bubble wall velocities in local equilibrium with the help of the novel constraint following from Eq.~\eqref{eq:gammaT},
\begin{align}
    \gamma_+T_+=\gamma_-T_-\, .\label{eq:conditionC}
\end{align}
This matching condition applies to local equilibrium only.

\subsection{Detonations}
\label{sec:detonations}

Before we study the bubble wall velocities in detail, it would be beneficial to see how $F_{\rm back}$ behaves with the bubble wall velocity given the relation~\eqref{eq:relation-v+v-}. For detonations, we have $v_+=v_w$ and $T_+=T_{\rm nuc}$, where $T_{\rm nuc}$ is the nucleation temperature and 
\begin{align}
    \omega_+=\frac{4}{3}a_+T_+^4=\frac{4}{3} a_+T_{\rm nuc}^4\, .
\end{align}
Substituting the above quantities into $c=\omega_+\gamma_+^2 v_+$, we obtain from Eq.~\eqref{eq:backforce2}
\begin{align}
    \frac{F_{\rm back}}{A}=
    \frac{4}{3}a_+T_{\rm nuc}^4 \gamma_w^2 v_w(v_w-v_-)\,.
\end{align}
From the above formula, one might think that the backreaction force scales as $\gamma_w^2$ and grows without bound. However, $v_w-v_-$ could decrease faster, and this is indeed the case, as we now show.

\begin{figure}[h]
    \centering
    \includegraphics[scale=0.82]{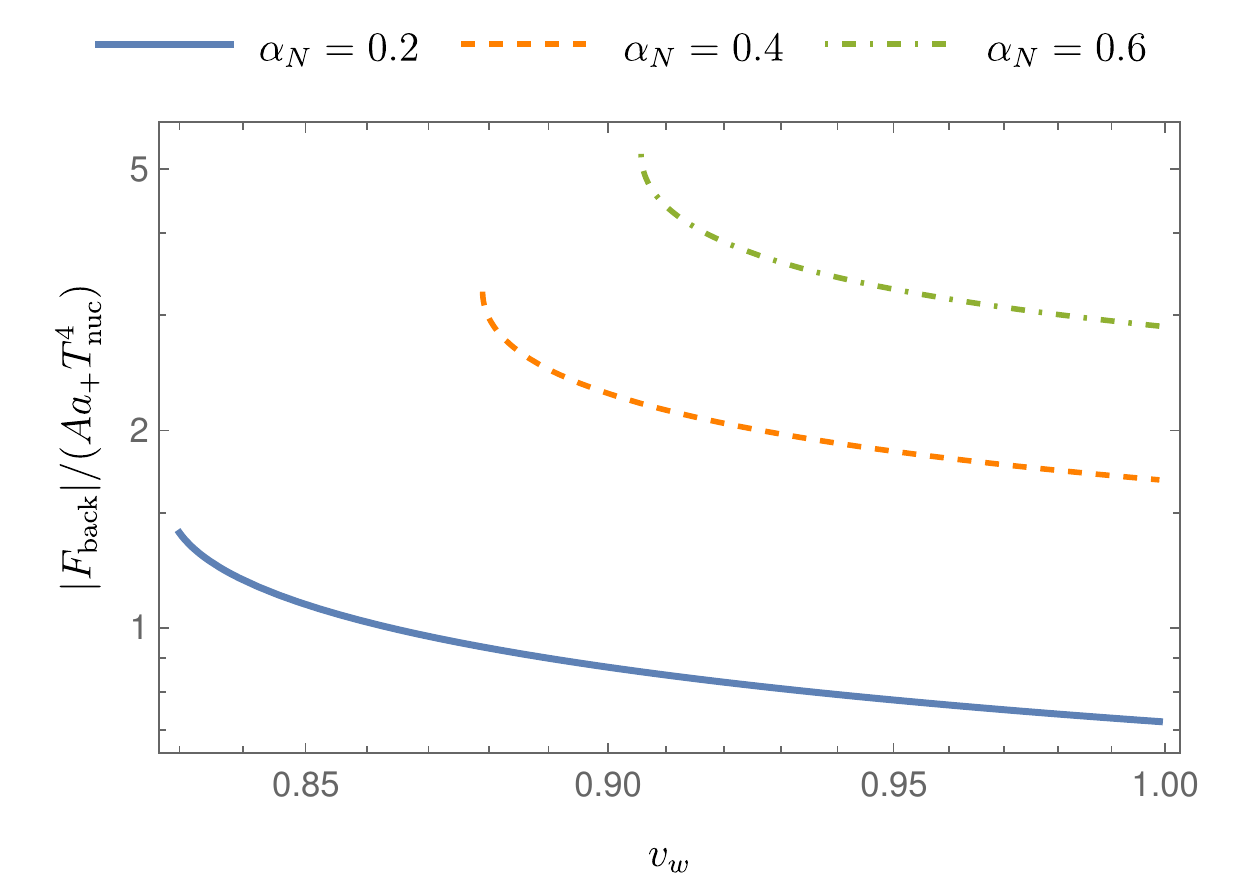}
    \caption{The function $f(v_w,\alpha_N)$ for given values of $\alpha_N$: $\alpha_N=0.2$ (thick), $\alpha_N=0.4$ (dot-dashed), $\alpha_N=0.6$ (dashed).}
    \label{fig:functionF}
\end{figure}

For detonations, $v_-$ belongs to the positive sign branch in Eq.~\eqref{eq:v-1}.\footnote{See Figure~1 in Ref.~\cite{Espinosa:2010hh}} Taking $v_+=v_w$ and $\alpha_+=\Delta\epsilon/(a(T_{\rm nuc})T^4_{\rm nuc})\equiv \alpha_N$, we obtain
\begin{align}
\label{eq:Fbackreaction}
    \frac{F_{\rm back}}{A}=a_+ T_{\rm nuc}^4 \,f(v_w,\alpha_N)\,,
\end{align}
where
\begin{align}
    f(v_w,\alpha_N)=&-\frac{2}{9} \gamma _w^2 \left[1-3 \alpha_N+3 \left(\alpha_N-1\right) v_w^2+\sqrt{\left(1-3 \alpha_N+3 \left(1+\alpha_N\right) v_w^2\right){}^2-12 v_w^2}\right]\,.
\end{align}
In Figure~\ref{fig:functionF}, we plot the function $f(v_w,\alpha_N)$ for various values of $\alpha_N$. It is clearly exhibited that the function does not grow quadratically with the Lorentz factor $\gamma_w$. Actually, due to the faster decrease of $v_w-v_-$, the function $f(v_w,\alpha_N)$ decreases when $v_w$ approaches one. We hope that this  may clarify the misunderstanding that the backreaction force grows quadratically with $\gamma_w$ as one may expect from Eq.~\eqref{eq:force} by assuming constant fluid velocity and temperature across the bubble wall. 

From Eq.~\eqref{eq:alpha&r2}, we have 
\begin{align}
    r=\frac{a_+T_{\rm nuc}^4}{a_-T_-^4}\,.
\end{align}
Substituting the above equation and Eq.~\eqref{eq:conditionC} into Eq.~\eqref{eq:v+/v-}, we have
\begin{align}
\label{eq:v-2}
    v_-=v_w\frac{\left(\frac{\gamma_w}{\gamma_-}\right)^4 b+3(1+\alpha_N)}{3\left(\frac{\gamma_w}{\gamma_-}\right)^4b+(1-3\alpha_N)}\,,
\end{align}
where
\begin{align}
\label{eq:b}
    b=\frac{a_-}{a_+}\,.
\end{align}
Combining Eq.~\eqref{eq:v-1} (with $v_+=v_w$) with Eq.~\eqref{eq:v-2}, one could obtain the bubble wall velocity. As $a_{+/-}$ depend on both the temperature and the field value, the quantity $b$ is rather model-dependent. Here for simplicity, we will simply assume a fixed value for $b$. In the Standard Model there is of course no first-order phase transition but one would get that $b$ is not far from $0.85$, which will be cited in our example numerical calculations below. In order to allow the transition to happen, we require the free energy difference $\Delta\mathcal{F}=p_--p_+$ to be positive at constant temperature $T_{\rm nuc}$. This gives $\alpha_+=\alpha_N>(1-b)/3=0.05$.

\begin{figure}[h]
    \centering
    \includegraphics[scale=0.82]{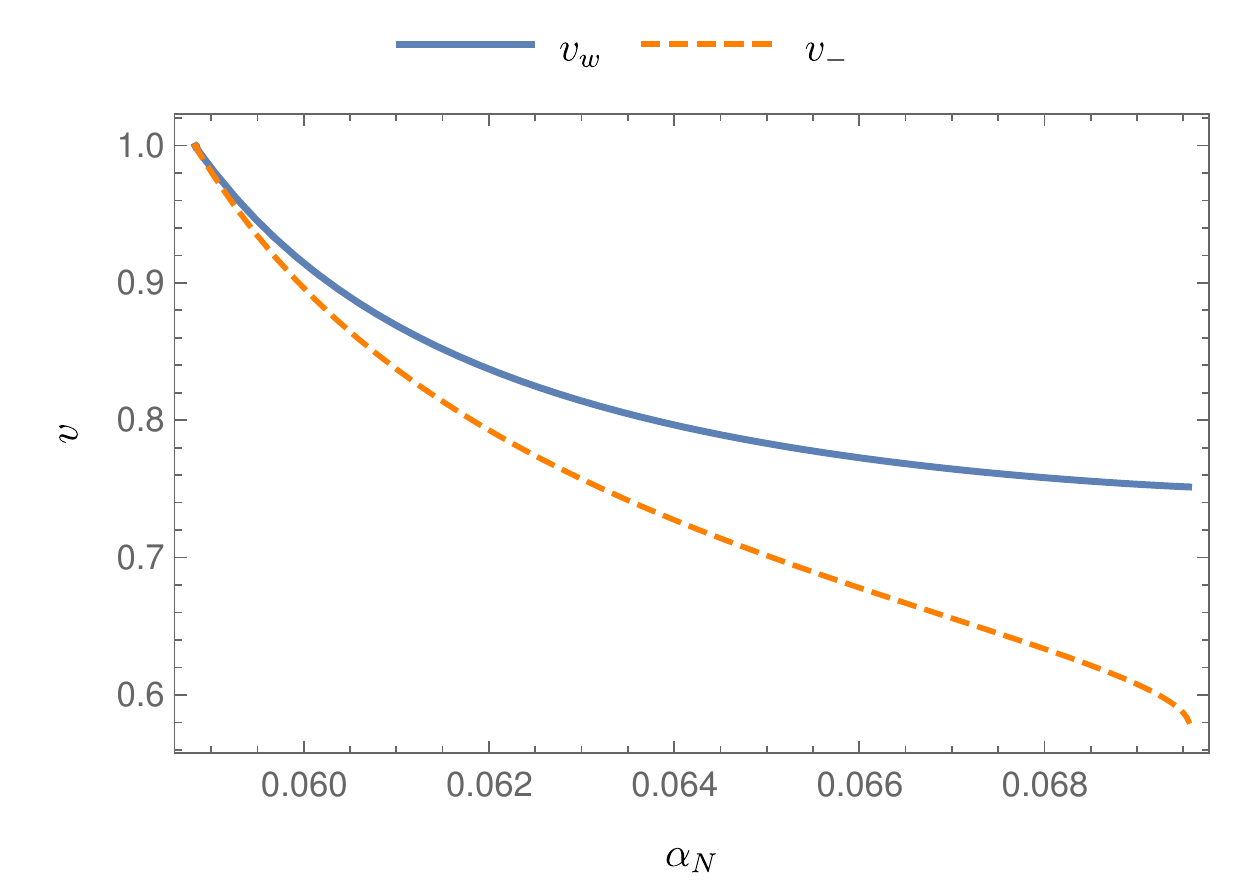}
    \caption{The bubble wall velocity $v_w$ and $v_-$ as a function of $\alpha_N$ in the detonation regime.}
    \label{fig:velocity-detonations}
\end{figure}

We numerically solve the coupled Eqs.~\eqref{eq:v-1} and~\eqref{eq:v-2} in the parameter region $\alpha_N\in(0.05,1]$. The behavior of the bubble wall velocity is shown in Figure~\ref{fig:velocity-detonations}. We find that solutions exist only in some region $\alpha_N\in[\alpha_{\rm min},\alpha_{\rm max}]$ with $\alpha_{\rm min}=0.0588236$, $\alpha_{\max}=0.0695655$ (these numbers of course depend on the working precision and maybe also the numerical methods used). The bubble wall velocity and $v_-$ decrease as $\alpha_N$ increases. The upper bound $\alpha_{\rm max}$ is due to the requirement of $v_-\geq 1/\sqrt{3}$. As $\alpha_N$ approaches $\alpha_{\rm max}$, $v_-$ approaches the sound speed $1/\sqrt{3}$ from above. For $\alpha_N>\alpha_{\rm max}$, the steady state satisfying all the three matching conditions does not exist in the detonation regime. This does not necessarily mean that the bubbles in this region of $\alpha_N$ must run away since the bubbles may not be able to enter the detonation regime at all. In the next subsection, we shall examine the bubble expansion in the deflagration regime in local equilibrium. Below we shall see that for deflagrations there is a solution for $\alpha_N\in [\alpha'_{\rm min},\alpha'_{\rm max}]$ where $\alpha'_{\rm min}$ is a numerical lower bound that is very close to $0.05$ and $\alpha'_{\rm max}$ is a number slightly larger than $\alpha_{\rm max}$. This would mean that only for $\alpha_N>\alpha'_{\rm max}$ the bubble wall will run away.

The velocities $v_w$ and $v_-$ approach the speed of light as $\alpha_N$ decreases. At $\alpha_N=\alpha_{\rm min}$, the velocities are very close to one, with $v_-=0.999997$, $v_w=0.999998$. One may therefore think that for sufficiently small $\alpha_N$ the bubbles essentially run away. However, in the next subsection, we shall see that in this region for $\alpha_N$, there are solutions with small $v_w$ in the deflagration regime. Therefore, the bubble wall does not run away for very small $\alpha_N$.   

\begin{figure}[h]
    \centering
    \includegraphics[scale=0.82]{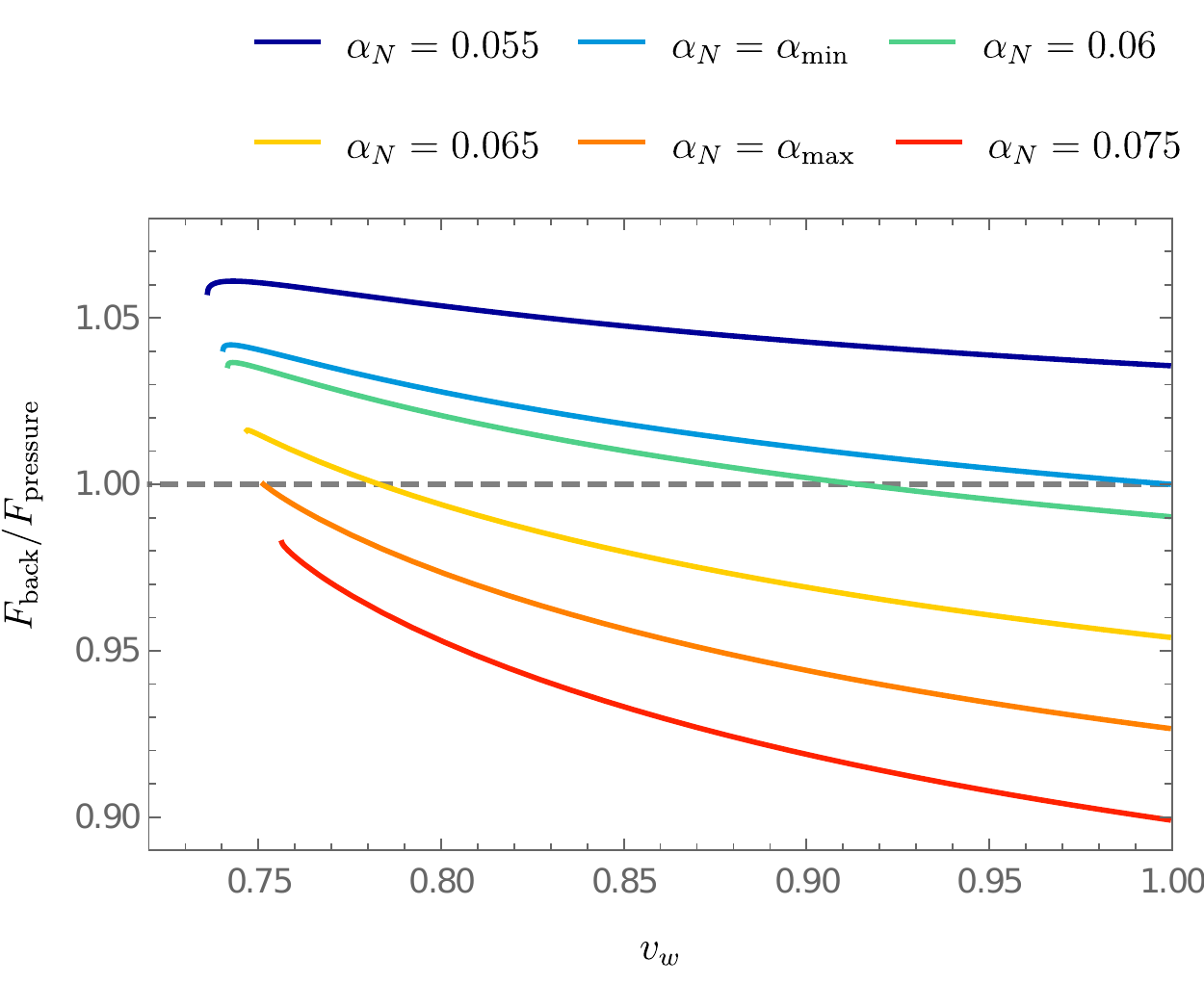}
    \caption{The ratio between $F_{\rm back}$ and $F_{\rm pressure}$ in the detonation regime as a function of the bubble wall velocity.}
    \label{fig:ratio}
\end{figure}

The behaviour of the velocities as a function of $\alpha_N$ shown in Figure~\ref{fig:velocity-detonations} may appear surprising compared to the case when dissipative friction is present, see, e.g., Figure~10 of Ref.~\cite{Espinosa:2010hh}. In the presence of the dissipative friction the velocities increase as $\alpha_N$ increases, while Figure~\ref{fig:velocity-detonations} shows the opposite behaviour. This discrepancy is a consequence of the difference in behaviour of the dissipative friction force and the non-dissipative backreaction force as a function of the bubble wall velocity. While the dissipative friction is an increasing function of the bubble wall velocity $v_w$, the backreaction force in local equilibrium is a decreasing function for detonations, see Figure~\ref{fig:functionF}. For a faster bubble wall, the dissipative friction is larger and needs to be balanced by a larger driving force, which requires a stronger phase transition. For the case of the non-dissipative backreaction force for detonations, with a larger bubble velocity the balance requires a smaller driving force and hence a weaker phase transition. More specifically, we can look into the ratio between $F_{\rm back}$ and $F_{\rm pressure}$. In the detonation regime, the former is given by 
\begin{align}\label{eq:Fpressure}
\frac{F_{\rm pressure}}{A}=-\Delta p=a_+ T_{\rm nuc}^4\left[-\frac{1}{3}+\frac{b}{3}\left(\frac{\gamma_w}{\gamma_-}\right)^4+\alpha_N\right]\,,
\end{align}
where we have used the matching condition~\eqref{eq:conditionC}, $T_+=T_{\rm nuc}$, $v_+=v_w$, and $\alpha_+=\alpha_N$. 
From Eqs.~\eqref{eq:Fbackreaction} and \eqref{eq:Fpressure} it follows that the ratio between $F_{\rm back}$ and $F_{\rm pressure}$ does not depend on $T_{\rm nuc}$. When making use of  Eq.~\eqref{eq:v-1}, $F_{\rm back}/F_{\rm pressure}$ becomes a function of only $v_w$ and $\alpha_N$, which is illustrated in Figure~\ref{fig:ratio}. One can see that for a value of $\alpha_N$ smaller than $\alpha_{\rm min}$, the backreaction force is always bigger than the driving force such that the balance between them can never be reached, while $\alpha_{\rm max}$ is another marginal point for the balance to exist.  In the region $(\alpha_{\rm min}, \alpha_{\rm max})$, the balance can be reached with a lower bubble velocity for a bigger value of $\alpha_N$. For $\alpha_N=\alpha_{\rm min}$, the balance can be reached for $v_w\approx 1$, as shown by the light blue line. 

To end this section, let us note that the family of detonation solutions in local equilibrium found here matches the behaviour of the branch of detonations with lower wall velocities found in Ref.~\cite{Megevand:2009ut}. As seen in Fig.~4 of the former reference, for low values of $\alpha_N$ the authors found wall velocities decreasing with $\alpha_N$. This branch of solutions was discarded on the grounds that it contains strong detonations with $v_-<c_s$, believed to be unphysical. However, the results in Ref.~\cite{Megevand:2009ut} show that not all of the solutions within this branch are strong detonations, and in fact the solutions having $\d v_w/\d\alpha_N<0$  have $v_-> c_s$ and are hence physical. These correspond to the detonations found in the present work. Note that for our solutions one has indeed $v_->c_s$, as illustrated in Fig.~\ref{fig:velocity-detonations}. The other branch of detonations in Ref.~\cite{Megevand:2009ut}, featuring $\d v_w/\d\alpha_N>0$ in agreement with the detonations of Ref.~\cite{Espinosa:2010hh}, does not seem to survive in local equilibrium, as the wall velocity grows for decreasing dissipative friction, in such a way that the bubbles become luminal before the dissipative force reaches zero. 

\subsection{Deflagrations}

For deflagrations, one has $v_-=v_w$. Taking the negative sign branch of Eq.~\eqref{eq:relation-v+v-}, we have
\begin{align}
\label{eq:v+2}
 v_+=\frac{1}{6 \left(1+\alpha _+\right) v_w}\left[1+3 v_w^2-\sqrt{1+6 \left(6 \alpha _+^2+4 \alpha _+-1\right) v_w^2+9 v_w^4}\right]\,.
\end{align}
Similarly, the second relation between $v_+$ and $v_w$ (from Eqs.~\eqref{eq:v+/v-} and~\eqref{eq:conditionC}) is given by
\begin{align}
\label{eq:v+3}
v_+=v_w\frac{3b+(1-3\alpha_+)\left(\frac{\gamma_w}{\gamma_+}\right)^4}{b+3(1+\alpha_+)\left(\frac{\gamma_w}{\gamma_+}\right)^4}\,.
\end{align}

As deflagrations feature a shock wave in front of the bubble wall, in general $T_+$ is not equal to $T_{\rm nuc}$ so that $\alpha_+\neq \alpha_N$. Therefore, one still needs to relate $\alpha_+$ to $\alpha_N$, i.e.,
\begin{align}
\label{eq:relation-alpha+alphaN}
\alpha_+=\frac{\Delta\epsilon}{a_+T_+^4}=\frac{\Delta\epsilon}{a_{\rm nuc}T_{\rm nuc}^4}\frac{a_{\rm nuc}T^4_{\rm nuc}}{a_+T^4_+}=\alpha_N\frac{q^4}{\tilde{b}}\,,
\end{align}
where we have defined $\tilde{b}=a_+/a_{\rm nuc}$---with $a_{\rm nuc}$ the value of the bag parameter $a$ in the symmetric phase at $T=T_{\rm nuc}$---and $q=T_{\rm nuc}/T_+$. The parameter $\tilde{b}$ is model dependent, similar to the parameter $b$. For a given model, $\tilde{b}$ can in principle be expressed in terms of $q$ and $T_{\rm nuc}$. For practical purposes, one can assume $\tilde{b}=1$. This is the case if either the temperature does not change much across the shock-wave front or the number of relativistic degrees of freedom is not sensitive to the temperature change. We will assume this value of $\tilde{b}$ in our numerical calculations below. 

To find the ratio between $T_{\rm nuc}$ and $T_+$, one has to solve the fluid velocity and temperature profile away from the bubble wall. The equations of motion are given by the continuity equations
\begin{subequations}
\label{eq:contituity}
\begin{align}
    &u_\nu \partial_\mu T^{\mu\nu}_f=0\,,\\
    &\Bar{u}_\nu\partial_\mu T_f^{\mu\nu}=0\,,
\end{align}
\end{subequations}
where $\Bar{u}^\mu$ is the normalized vector orthogonal to $u^\mu$. It is convenient now to work in the rest frame of the bubble center. Then we have $u^\mu=\gamma(1,\vec{v})$ and $\Bar{u}^\mu=\gamma(v,\vec{v}/v)$. In spherical coordinates, $u^\mu=(\gamma,\gamma v,0,0)$ and $\Bar{u}^\mu=(\gamma v,\gamma,0,0)$. Since there is no characteristic scale in the problem, one will obtain a similarity solution that depends only on the dimensionless variable $\xi=r/t$~\cite{Gyulassy:1983rq}. From Eqs.~\eqref{eq:contituity}, one has 
\begin{subequations}
\begin{align}
   & (\xi-v)\frac{\partial_\xi\rho}{\omega}=2\frac{v}{\xi}+[1-\gamma^2v(\xi-v)]\partial_\xi v\,,\\
   & (1-v\xi)\frac{\partial_\xi p}{\omega}=\gamma^2(\xi-v)\partial_\xi v\,. 
\end{align}
\end{subequations}
These two equations can be rearranged into 
\begin{subequations}
\label{eq:continuity2}
\begin{align}
    &2\frac{v}{\xi}=\gamma^2(1-v\xi)\left[\frac{\mu^2}{c_s^2}-1\right]\partial_\xi v\,,\\
    &\partial_\xi\omega=\omega\left(1+\frac{1}{c_s^2}\right)\gamma^2\mu\partial_\xi v\,,
\end{align}
\end{subequations}
where 
\begin{align}
    c^2_s=\,\frac{\partial_T p}{\partial_T \rho}\,,\qquad \mu=\,\frac{\xi-v}{1-\xi v}\,.
\end{align}
Solving these equations relies on numerical calculations. Dramatic simplifications can be obtained in the planar approximation under which the term with $1/\xi$ can be dropped. In this case, the fluid velocity and temperature in the shock wave are constant. These quantities have a discontinuity at the shock-wave front. Below, we shall use the planar approximation.

At the shock-wave front, one has similar matching conditions for the fluid velocity and temperature. We will denote the various quantities defined in the frame of the shock-wave front with a tilde. Since the value of the background scalar field does not change across the shock-wave front, we have $\tilde{\alpha}_+=0$. The corresponding relations analogous to Eqs.~\eqref{eq:conditionA},~\eqref{eq:conditionB} now give 
\begin{subequations}
\begin{align}
\label{eq:tildev+}
\tilde{v}_+&=v_{\rm front}=\sqrt{\frac{3+\tilde{r}}{3(1+3\tilde{r})}}\,,\\
\label{eq:tildev-}
 \tilde{v}_-&=\frac{1}{3\tilde{v}_+}=\frac{1}{3v_{\rm front}}\,,
\end{align}
\end{subequations}
where
\begin{align}
\tilde{r}=\frac{\tilde{a}_+ \widetilde{T}^4_+}{\tilde{a}_-\widetilde{T}^4_-}=\left(\frac{\tilde{a}_+}{\tilde{a}_-}\right)\left(\frac{T_{\rm nuc}}{T_+}\right)^4\equiv \frac{q^4}{\tilde{b}}\,.
\end{align}
Here we have used the relations $\widetilde{T}_+=T_{\rm nuc}$, $\widetilde{T}_-=T_+$, $\tilde{a}_+=a_{\rm nuc}$ and $\tilde{a}_-=a_+$. The last relation is due to the fact that the temperature and the background scalar field do not change in the region between the bubble wall and the shock-wave front. On the other hand, $\tilde{v}_-$ can be related to $v_+$, $v_w$ and $v_{\rm front}$ by considering two relations for the fluid velocity,
\begin{subequations}
\begin{align}
v_{\rm fluid}&=-\frac{v_+-v_-}{1-v_+v_-}=-\frac{v_+-v_w}{1-v_+v_w}\,,\\
v_{\rm fluid}&=-\frac{\tilde{v}_--\tilde{v}_+}{1-\tilde{v}_+\tilde{v}_-}=-\frac{\tilde{v}_--v_{\rm front}}{1-v_{\rm front}\tilde{v}_-}\,.
\end{align}
\end{subequations}
Combining these two equations gives 
\begin{align}
\label{eq:tildev-2}
 \tilde{v}_-=\frac{v_+-v_w+v_{\rm front}-v_{\rm front}v_+v_w}{1+v_+v_{\rm front}-v_+v_w-v_w v_{\rm front}}\,.
\end{align}
Equating Eq.~\eqref{eq:tildev-} to the above equation gives
\begin{align}
\label{eq:v-=v-}
\frac{v_+-v_w+v_{\rm front}-v_{\rm front}v_+v_w}{1+v_+v_{\rm front}-v_+v_w-v_w v_{\rm front}}=\frac{1}{3 v_{\rm front}}\,.
\end{align}

Equations~\eqref{eq:v+2},~\eqref{eq:v+3} (after substituting~\eqref{eq:relation-alpha+alphaN}) and Eq.~\eqref{eq:v-=v-} (after substituting~\eqref{eq:tildev+})  provide three equations for the three unknown quantities $(v_w,v_+,q)$ (recall that we take $\tilde{b}=1$). Note that at the shock front local entropy conservation is not satisfied. This can be seen from the following incompatibility between the condition\footnote{We thank the referee for pointing this out.} 
\begin{align}
\label{eq:tildeTtildegamma}
\widetilde{T}_-\tilde{\gamma}_-=\widetilde{T}_+\tilde{\gamma}_+\quad \Rightarrow\quad T_+\tilde{\gamma}_-=T_{\rm nuc}\tilde{\gamma}_+\,,
\end{align}
and Eqs.~\eqref{eq:tildev+},~\eqref{eq:tildev-}. From Eqs.~\eqref{eq:tildev+},~\eqref{eq:tildev-}, we have 
\begin{align}
q=\left(\frac{3(1-v^2_{\rm front})}{9v^2_{\rm front}-1}\right)^{1/4}\,,
\end{align}
while Eq.~\eqref{eq:tildeTtildegamma} and Eq.~\eqref{eq:tildev-} give
\begin{align}
q=\frac{\tilde{\gamma}_-}{\tilde{\gamma}_+}=\left(\frac{9v^2_{\rm front}(1-v^2_{\rm front})}{9 v^2_{\rm front}-1}\right)^{1/2}\,.
\end{align}
We  note that the lack of entropy conservation at the shock front can be avoided by going beyond the planar wall approximation, and solving the hydrodynamic equations for the fluid in front of the bubble. In this case one can get solutions which contain no discontinuities in the velocity and entropy, see e.g.~Ref.~\cite{Balaji:2020yrx}.

\begin{figure}[h]
    \centering
    \includegraphics[scale=0.82]{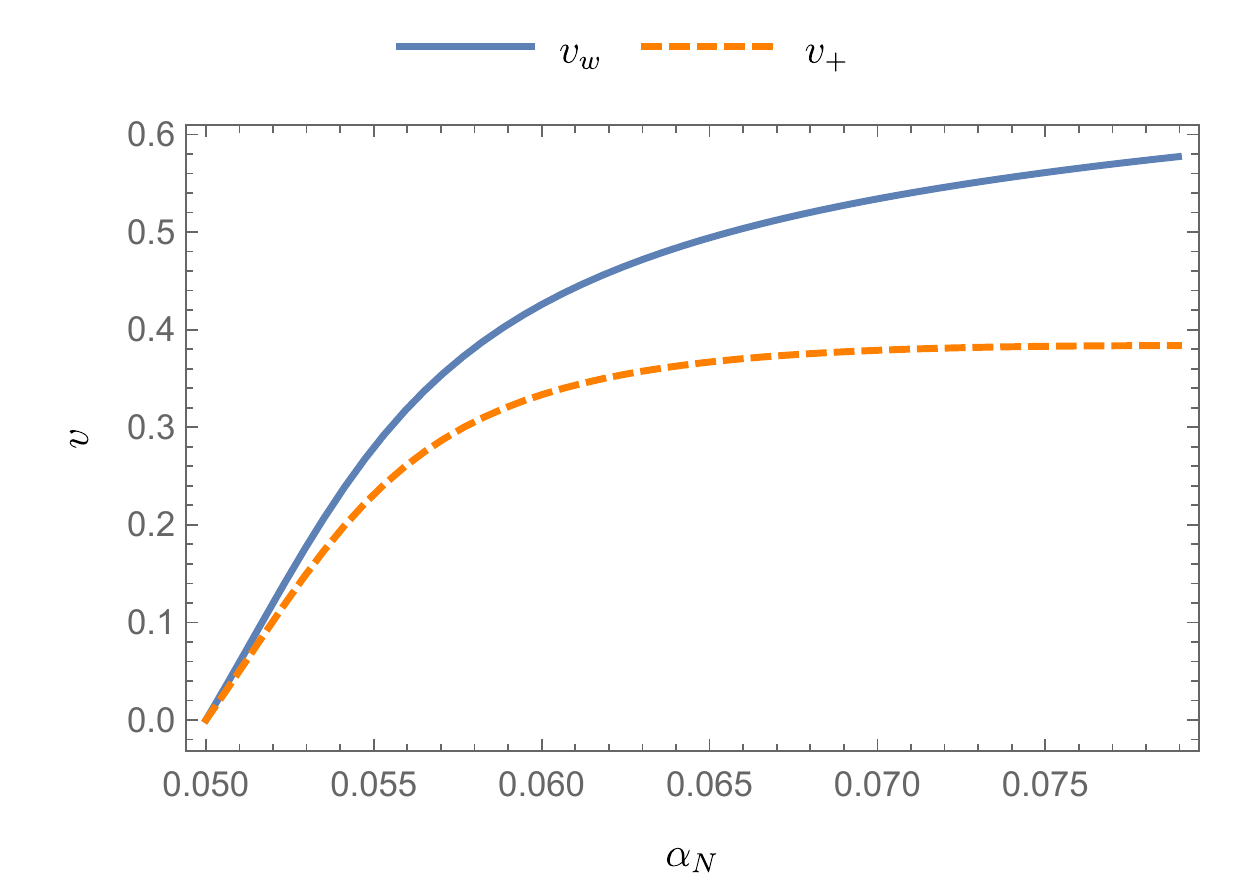}
    \caption{The bubble wall velocity $v_w$ and $v_+$ as a function of $\alpha_N$ in the deflagration regime.}
    \label{fig:velocity-deflagrations}
\end{figure}

\begin{figure}[h]
    \centering
    \includegraphics[scale=0.82]{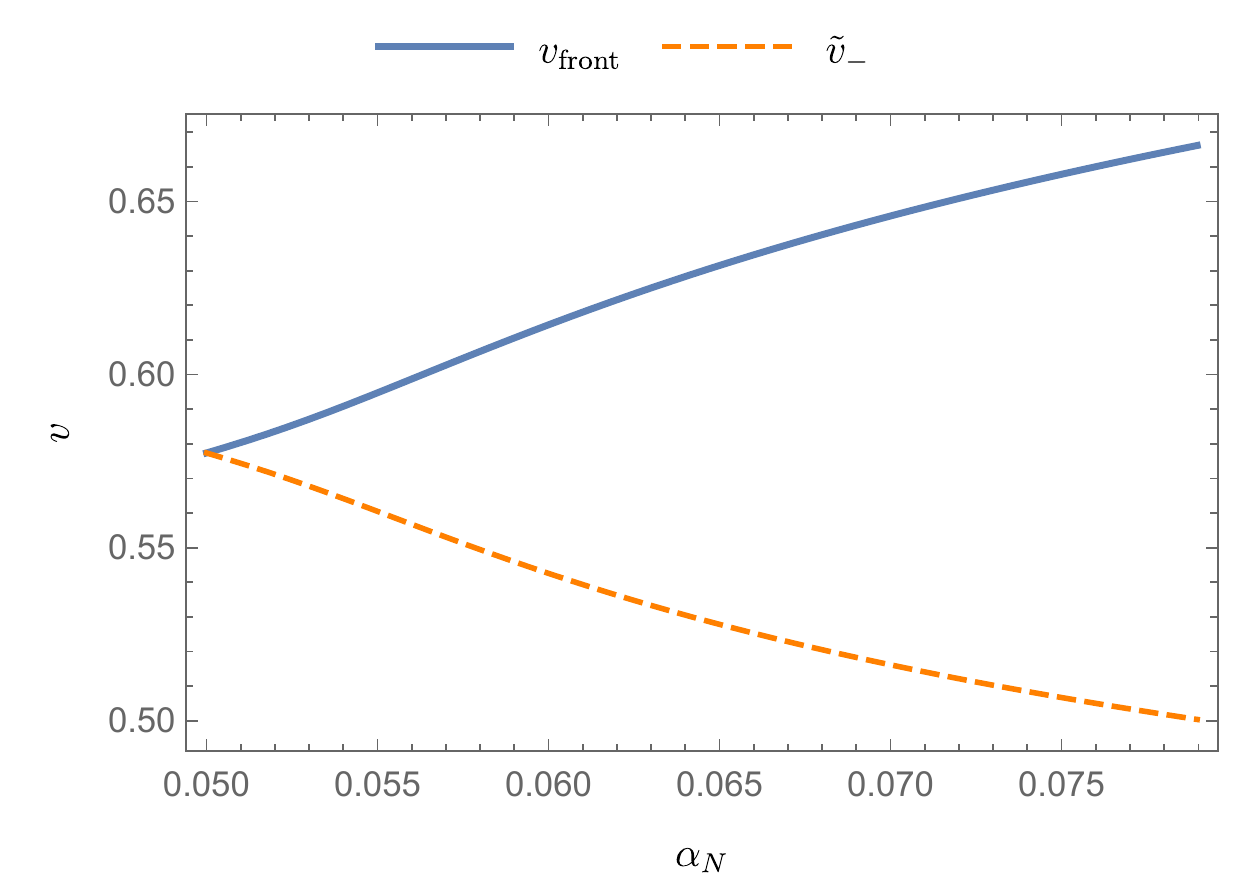}
    \caption{The front velocity $v_{\rm front}$ and $\tilde{v}_-$ as a function of $\alpha_N$ in the deflagration regime.}
    \label{fig:vfront}
\end{figure}

We solve the coupled equations numerically. The behavior of the velocities is shown in Figure~\ref{fig:velocity-deflagrations}. Similar to the case of the detonation regime, we find that the solution exists only for a parameter region $\alpha_N\in[\alpha'_{\rm min},\alpha'_{\rm max}]$ with $\alpha'_{\rm min}=0.0500061$, $\alpha'_{\rm max}=0.0789645$.  The lower bound $\alpha'_{\rm min}$ is very close to the static limit of the lower bound for the phase transition itself (see the comment below Eq.~\eqref{eq:b}) such that one cannot necessarily distinguish them. The velocities increase as $\alpha_N$ increases. The bubble wall velocity approaches the sound speed from below as $\alpha_N$ approaches $\alpha'_{\rm max}$, while it approaches zero as $\alpha_N$ approaches $\alpha'_{\rm min}$. For $\alpha_N>\alpha'_{\rm max}$, there are solutions neither in the detonation nor the deflagration regime. Thus, the bubble wall in this parameter region runs away. We also show the front velocity $v_{\rm front}$ and $\tilde{v}_-$ in Figure~\ref{fig:vfront} as a function of $\alpha_N$ in the region where the solution exists. It can be seen that $v_{\rm front}>\tilde{v}_-$, as expected.

\begin{figure}[h]
    \centering
    \includegraphics[scale=0.82]{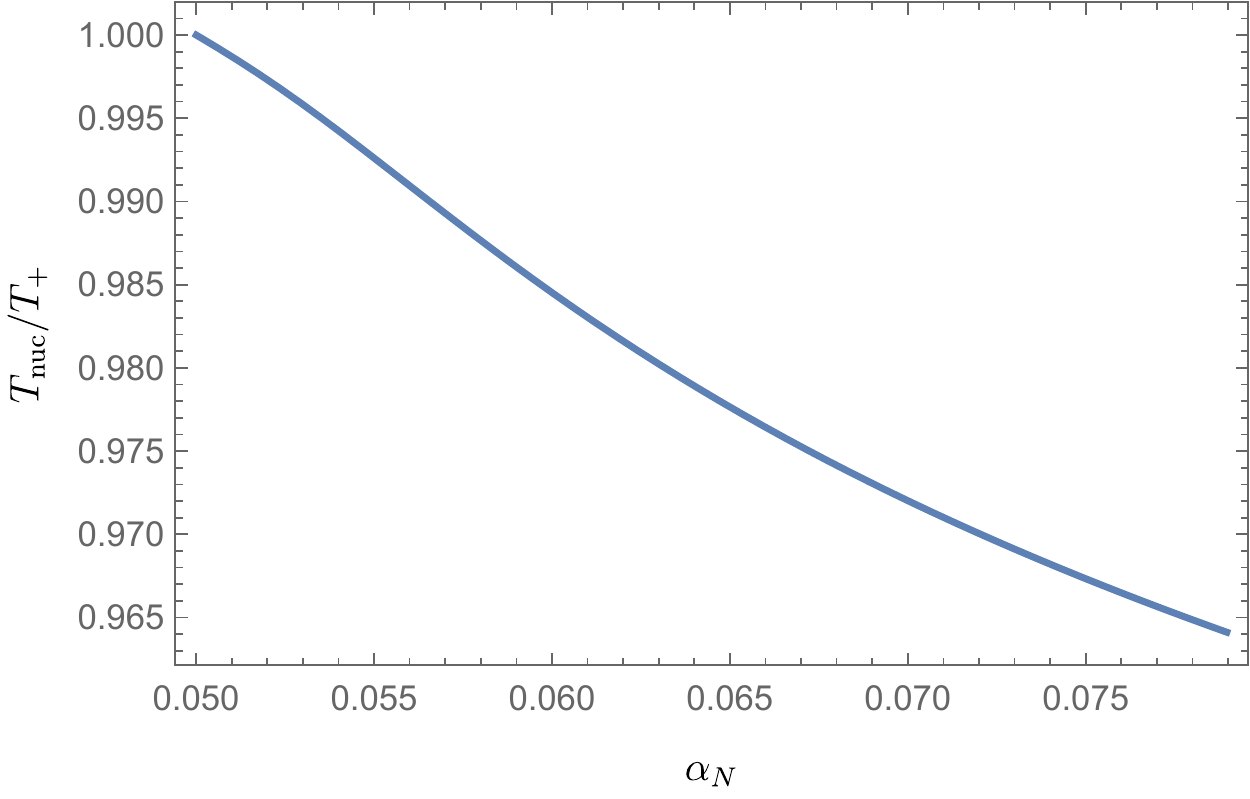}
    \caption{The ratio between the nucleation temperature and the temperature in front of the bubble wall, $T_{\rm nuc}/T_+$, in the deflagration regime as a function of the phase transition strength $\alpha_N$.}
    \label{fig:q}
\end{figure}

In Figure~\ref{fig:q} we show the ratio between the nucleation temperature, $T_{\rm nuc}$, and the temperature in front of the bubble wall, $T_+$, as a function of the phase transition strength. We see that $T_{\rm nuc}$ is always smaller than $T_+$, in agreement with Ref.~\cite{Espinosa:2010hh}. The ratio $T_{\rm nuc}/T_+$ decreases as $\alpha_N$ increases. Note the minimum value for the ratio $T_{\rm nuc}/T_+\approx 0.86$ at $\alpha_N=\alpha'_{\rm max}$, which again justifies the assumption of $\tilde{b}=1$. 

\begin{figure}[h]
    \centering
    \includegraphics[scale=0.82]{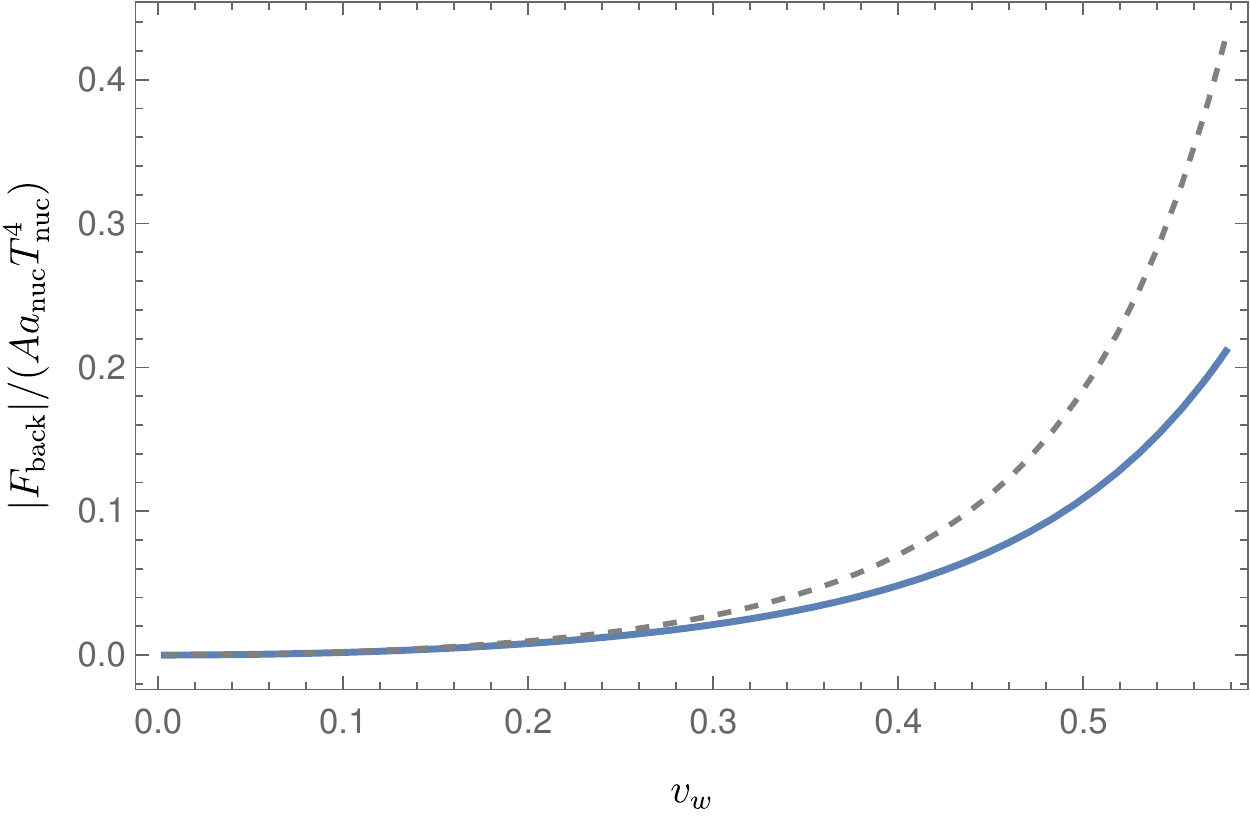}
    \caption{Normalized backreaction force as a function of the wall velocity for deflagrations, using the exact expression (solid blue) and the approximation of Eq.~\eqref{eq:force}  (dashed gray). }
    \label{fig:Fback}
\end{figure}

Given values of $v_w,v_+,q$ satisfying all the constraints, one can  recover the backreaction force of Eqs.~\eqref{eq:inte}, \eqref{eq:backforce2} up to a factor of $a_{\rm nuc}T_{\rm nuc}^4$ as
\begin{align}
 \frac{F_{\rm back}}{A }=\frac{4}{3}\,a_+ T_+^4 \gamma_+^2v_+(v_+-v_w)=\frac{4}{3 }a_{\rm nuc} T_{\rm nuc}^4\frac{\tilde{b}}{q^4}\gamma_+^2v_+(v_+-v_w)\,.
\end{align}
As shown in Figure~\ref{fig:Fback}, the resulting force now grows with the velocity but, as was noted in  Ref.~\cite{Balaji:2020yrx}, differs from the approximate result of Eq.~\eqref{eq:force}, which overestimates the force by a factor between 1.2 and 2.

To conclude, let us comment that the fact that  solutions to  Eqs.~\eqref{eq:v+2},~\eqref{eq:v+3} and~\eqref{eq:v-=v-} exist for the parameter region $[\alpha'_{\rm min},\alpha'_{\rm max}]$ with $\alpha'_{\rm max}>\alpha_{\rm max}$ indicates the following picture. In local equilibrium and under the assumptions that we have used, there is a critical value for the transition strength, $\alpha_{\rm crit}=\alpha'_{\rm max}$. For $\alpha_N<\alpha_{\rm crit}$, the bubble wall does not run away while for $\alpha_N>\alpha_{\rm crit}$, it does.

\section{Matching to particle physics models}
\label{sec:microphysics}

As mentioned earlier, when admitting a background and temperature dependence of the parameters $a,\epsilon$ in the bag model of Eq.~\eqref{eq:bagprho}, one can capture arbitrary dispersion relations. This allows to map the results of the previous section to microscopic particle physics models. 

In a given particle physics model, the total pressure $p$ corresponds to minus the total effective potential $V_{\rm eff}(\phi,T)$ (see Eq.~\eqref{eq:P}) which can be computed using standard methods. Once $p$ is known, well-known thermodynamical identities then allow to calculate the total density $\rho$ as
\begin{align}\label{eq:rhofromp}
\rho(\phi,T)= T\frac{\partial p(\phi,T)}{\partial T}-p(\phi,T).
\end{align}
With this, for a constant background $\phi$ one can identify the bag parameters in Eq.~\eqref{eq:bagprho} as
\begin{align}\label{eq:aeps}\begin{aligned}
 a(\phi,T)=&\,\frac{3}{4 T^4}\,(\rho(\phi,T)+p(\phi,T))\,,\\
 \epsilon(\phi,T)=&\,\frac{1}{4}\,(\rho(\phi,T)-3p(\phi,T))\,.
\end{aligned}\end{align}
In the discussion of the previous section, it was concluded that in local equilibrium the knowledge of
the bag parameters $a,\epsilon$ in the two phases separated by the bubble wall allows to infer the
wall velocity. In practice, as $a,\epsilon$ depend on the scalar background, their computation requires to 
solve the scalar profile across the wall. In the wall frame and in local equilibrium, assuming a stationary 
expansion and a planar limit, the scalar field
satisfies the static limit of Eq.~\eqref{eq:eomdeltaf} with the non-equilibrium contributions omitted,
\begin{align}\label{eq:scalareom}
 -\frac{\partial^2}{\partial z^2}\phi+\frac{\partial V_{\rm eff}(\phi,T)}{\partial\phi}=0\,.
\end{align}
The boundary conditions must enforce that the field is in the symmetric phase
$\phi=0$ (or the false-vacuum phase for more general cases) in front of the bubble, as well as zero first and second derivatives
for $z\rightarrow\pm\infty$. The condition of zero second derivatives ensures that the field
stabilizes into a minimum of the potential far away from the bubble, as is clear
from Eq.~\eqref{eq:scalareom}. As the scalar equation of motion involves the temperature, one needs to solve
Eqs.~\eqref{eq:em-con-1} and \eqref{eq:em-con-2} for the temperature profile across the wall. As done in Refs.
~\cite{Ignatius:1993qn} and \cite{Balaji:2020yrx}, the equations can be used to express $v,T$ in terms 
of $\phi,\phi'=\d\phi/\d z$.
When substituting $T(\phi,\phi')$ into Eq.~\eqref{eq:scalareom} one gets a single differential equation for $\phi$ 
with a modified potential. Once the solution for the scalar field is found, the profiles for $v$ and $T$ 
immediately follow. A
subtlety emphasized in Ref.~\cite{Balaji:2020yrx} is that the solutions $v(\phi,\phi'), T(\phi,\phi')$
can be multi-valued, and accounting for this allows to find the first explicit examples of static solutions for the scalar and
hydrodynamic profiles for detonations in local equilibrium. In the same reference it was noted that 
the resulting backreaction force had order one deviations with respect to Eq.~\eqref{eq:force}, 
yet the dependence of the force with respect to the velocity was not studied. For illustration, here we will 
show a family of detonation solutions obtained with the above procedure and for which the backreaction
force decreases with the wall velocity as expected from the analysis of Section~\ref{sec:detonations}. 

For this we choose as in Ref.~\cite{Balaji:2020yrx} an extension of the Standard Model by a multiplet $\varphi$ of $N$ 
complex singlets, which allows for first-order transitions for the Higgs doublet $H$. Assuming a $U(N)$ symmetry 
involving the singlet fields, the Lagrangian can be taken as
\begin{eqnarray}
{\cal L} \supset -m^2_{H}H^\dagger H-\frac{\lambda}{2}\,(H^\dagger H)^2-m^2_{\varphi}\,\varphi^\dagger\varphi
-\frac{\lambda_\varphi}{2}\, (\varphi^\dagger\varphi)^2-\lambda_{H\varphi}H^\dagger H \varphi^\dagger\varphi \; .
\end{eqnarray}
Assuming a background for the neutral component $\phi=\sqrt{2}\,{\rm Re}\,H^0$ of the Higgs, within a high temperature expansion one can compute the total pressure as
\begin{align}\label{eq:pressuremodel}
 \begin{aligned}
 &p(\phi,T)=\,\frac{\pi ^2 T^4}{90}  (g_{*,\rm SM}+2N)-T^2\left(\phi^2 \left(\frac{y_b^2}{8}+\frac{3 g_1^2}{160}+\frac{3 g_2^2}{32}+\frac{\lambda }{8}+\frac{N \lambda _{H\varphi}}{24}+\frac{y_t^2}{8}\right)+\frac{m^2_H}{6}+\frac{N m^2_\varphi}{12}\right)\\
 &-\frac{T}{12\pi}\left(-\frac{3}{4} \left(g_2 \phi\right)^3-\frac{3 \phi^3}{8} \left(\frac{3 g_1^2}{5}+g_2^2\right)^{3/2}\right.
 -3 \left(\frac{\phi^2 \lambda }{2}+m^2_H\right)^{3/2}-\left(\frac{3 \phi^2 \lambda }{2}+m^2_H\right)^{3/2}\\
 &\left.-2 N \left(\frac{\phi^2 \lambda _{H\varphi}}{2}+m^2_\varphi\right)^{3/2}\right)+\frac{m^2_H}{2}\phi^2-\frac{\lambda}{8}\,\phi^4\,.
  \end{aligned}
\end{align}
In the previous expression $\phi$ designates the neutral Higgs field, $g_1$ and $g_2$ are the hypercharge and weak gauge couplings, while  $y_b, y_t$  are the bottom and top quark Yukawa couplings. For the gauge couplings we use the normalization compatible with Grand Unification. The number of effective relativistic degrees of freedom in the SM plasma is denoted as $g_{\star,\rm SM}\sim 106.75$. We fix $N=1$, $m^2_S=400 ({\rm GeV})^2$, $\lambda_\varphi=0.085$, and vary $\lambda_{H\varphi}$.  As usual we estimate the nucleation temperature by solving for critical configurations of the static action with three-dimensional spherical symmetry, assuming a constant temperature, and demanding $S_3[\phi_{\rm nuc}(r),T_{\rm nuc}]/T_{\rm nuc}\sim 140$ (see e.g. Ref.~\cite{Quiros:1999jp}). For the subsequent determination of the detonation profiles, we assume a planar rather than spherical symmetry, in order to make direct contact with the previous discussions. For different values of $\lambda_{H\varphi}$, Table~\ref{tab:examples}
gives the corresponding bag parameters, wall velocity, and backreaction force inferred from the numerical solutions. The bag parameters on each side of the wall were recovered by applying Eqs.~\eqref{eq:aeps},~\eqref{eq:pressuremodel} and \eqref{eq:rhofromp} at large $|z|$, when the background becomes approximately constant. The wall velocity $v_+$ corresponds to the fluid velocity at large $z$, while the backreaction force was obtained by applying Eqs.~\eqref{eq:inte} using $\omega=\rho+p$. As can be seen, the backreaction force decreases as the wall velocity increases, in accordance with the expectations from Section~\ref{sec:detonations}. As a cross-check of these numerical results, one can input the bag parameters into the equations of Section~\ref{sec:detonations} and check whether one predicts the same wall velocities. We have found that the same wall velocities can be obtained by adjusting the values of $\alpha_N$ resulting from Table~\ref{tab:examples} and Eq.~\eqref{eq:alpha&r2} to within less than one part in a thousand. These very small differences arise from the numerical uncertainty inherent in the shooting method employed for finding the scalar profiles within a finite interval of $z$. Within this method,  the derivatives with respect to $z$ of the scalar field do not vanish completely for large $z$, in contrast to what is assumed in Section~\ref{sec:detonations}.

\begin{table}[t]
\begin{center}\small
\begin{tabular}{c||c|c|c|c|c|c|c} 
$\lambda_{H\varphi}$ & $T_{\rm nuc}$ (GeV) & $a_+$ & $a_-$ & $\epsilon_+/m_W^4$  & $\epsilon_-/m_W^4$ & $v_w$ & $|F_{\rm back}|/A/m_W^4$
 \\[5pt]
 \hline
 \hline
 \rule{0pt}{17pt}
 0.75 & 135.1859 & 35.8790 & 35.8595 & $-0.2719$ & $-0.3238$ & 0.7601 & 0.1639
\\[5pt]
\hline
\rule{0pt}{17pt}
  0.80 & 134.6580 & 35.8798 & 35.8573 & -0.2697 & -0.3289 & 0.7559 & 0.1894
\\[5pt]
\hline
\rule{0pt}{17pt}
0.85 & 134.1440 & 35.8806 & 35.8546 & -0.2677 & -0.3349 & 0.7502 & 0.2200
\\[5pt]
\hline
\rule{0pt}{17pt}
0.90 & 133.6444 & 35.8813 & 35.8514 & -0.2657 & -0.3421 & 0.7436 & 0.2567
\\[5pt]
\hline
\rule{0pt}{17pt}
0.95 & 133.1594 & 35.8821 & 35.8477 & -0.2638 & -0.3503 & 0.7365 & 0.2997
\\[5pt]
 \hline
\end{tabular}
\end{center}
\caption{\label{tab:examples}Detonation configurations obtained by solving the static equations for the scalar field and the plasma in the static limit. Some quantities are given normalized by the mass of the $W$ boson, $m_W=80.379$ GeV.}

\end{table}

\section{Conclusions}
\label{sec:Conc}

In this work, we have carried out a systematic study of the bubble wall dynamics in local equilibrium, with the aim of improving the understanding of the origin of the friction force highlighted in Ref.~\cite{Mancha:2020fzw} and further studied in Ref.~\cite{Balaji:2020yrx}, where it was linked to the hydrodynamic obstruction of Ref.~\cite{Konstandin:2010dm}. A further goal of this paper has been to clarify whether this backreaction force really rules out runaway behaviors, as proposed in Ref.~\cite{Mancha:2020fzw} based on the proportionality of the force to $\gamma_w^2-1$. We have first provided a general analysis on steady bubble walls, identifying two different forces on the wall, namely the dissipative friction due to out-of-equilibrium effects and the non-dissipative backreaction force due to non-constant temperature distributions of the fluid across the bubble wall, as shown in Eq.~\eqref{eq:Ffr}. The latter is precisely the effective ``friction'' in local equilibrium of Ref.~\cite{Mancha:2020fzw}. We have further shown that a nonzero friction force in local equilibrium necessarily requires a temperature gradient across the wall, in contrast to the constant temperature assumption of Ref.~\cite{Mancha:2020fzw}. As a consequence, the backreaction force deviates from the $\gamma_w^2$-scaling proposed in the former reference. In local equilibrium the entropy is conserved locally (as follows from energy-momentum conservation and the equation of motion of the background) from which we find a new matching condition across the bubble wall, $\gamma_+T_+=\gamma_-T_-$. This new condition makes it possible to solve for the bubble velocity algebraically in terms of the bag parameters. We have constructed a closed system of equations for both the detonation and deflagration regimes and solved them numerically. We find that for both detonations and deflagrations, steady-state solutions exist only for a narrow parameter region. The numerical results indicate that there is a critical value of the phase transition strength, $\alpha_{\rm crit}$, below which the bubble wall does not run away and above which the bubble wall does run away. The deviation from the expectations of Ref.~\cite{Mancha:2020fzw} can be understood from the fact that the backreaction force is not strictly proportional $\gamma_w^2-1$. Although for deflagrations the force still grows with velocity, it remains below the expectations of Ref.~\cite{Mancha:2020fzw}, while for detonations the force actually decreases with the wall velocity.

\section*{Acknowledgments}

WYA is grateful to Yan-Bing Wei for the computation resources used in the numerical calculations.  CT acknowledges financial support by the DFG through the ORIGINS cluster of excellence.

\begin{appendix}
\renewcommand{\theequation}{\Alph{section}\arabic{equation}}
\setcounter{equation}{0}

\end{appendix}

\bibliographystyle{utphys}
\bibliography{FVDref}{}

\end{document}